\documentclass[11pt,letterpaper]{article}

\usepackage[margin=1.45in,top=1.3in,bottom=1.3in]{geometry}
\usepackage{mathptmx}
\usepackage{amsmath,amssymb,amsfonts}
\usepackage{amsthm}
\usepackage{graphicx}
\usepackage{xcolor}
\usepackage{authblk}
\usepackage{caption}
\usepackage{cite}
\usepackage{url}
\usepackage{microtype}
\usepackage[hidelinks]{hyperref}

\usepackage{pifont}

\usepackage{float}

\renewcommand{\abstractname}{Abstract}
\newtheorem{theorem}{Theorem}

\newtheorem{conjecture}{Conjecture}
\theoremstyle{definition}
\newtheorem{assumption}{Assumption}
\theoremstyle{remark}

\newenvironment{IEEEkeywords}%
  {\par\vspace{.6\baselineskip}\noindent\small\textbf{Index Terms---}\ignorespaces}%
  {\par\vspace{.4\baselineskip}}

\newcommand{\widewidth}{\textwidth}

\newcommand{\Obs}{A}                      
\newcommand{\Pre}{\mathcal{D}}            
\newcommand{\Xh}{\widetilde{X}}           
\newcommand{\Sh}{\widetilde{S}}           
\newcommand{\Kw}{K(W)}                    
\newcommand{\Rw}{R(W)}                    
\newcommand{\Xalph}{\Lambda}              
\newcommand{\Salph}{\mathcal{S}}          

\title{\bfseries Condensing Law of Security Information}

\author[1]{Qi Tan$^{*,}$}
\author[2]{Tianyu Cui\thanks{These authors contributed equally to this work.}$^,$}
\author[3]{Qi Li\textsuperscript{\ding{41}}$^,$}
\author[4]{Ke Xu\textsuperscript{\ding{41}}$^,$}
\affil[1]{College of Computer Science and Software Engineering, Shenzhen University}
\affil[2]{Independent researcher}
\affil[3]{Institute for Network Sciences and Cyberspace, Tsinghua University}
\affil[4]{Department of Computer Science and Technology, Tsinghua University}
\date{}

\begin{document}
\maketitle
\begin{abstract}
\noindent
Scaling laws indicate that the generalization error of a large language model (LLM) decreases monotonically with model size once data is sufficient, and the corollary that larger models outperform smaller ones is now widely accepted. We argue that this corollary fails in security: decisions there hinge on anomalous inputs, which are long-tailed (e.g., zero-day attacks) and therefore poorly served by likelihood fitting. Worse still, the capability brought about by scale (e.g., inferring the intended whole from a corrupted part) leads a LLM silently \emph{repair} the anomaly before any downstream security head ever sees it. We formalize this as a tradeoff between two quantities: the cognitive capability $\Kw=I(W;P)$ that a model holds about the world distribution, and the security-information retention $\Rw=\frac{I(Z;S)}{I(\Obs;S)}$ that survives in its intermediate representation. We derive three information-theoretic bounds to connect them: an upper bound on restoration accuracy in terms of $\Kw$, an upper bound on $\Rw$ in terms of the restoration error, and a Fano lower bound on security-task error in terms of $\Rw$. These theorems approximately imply that usable security information is a product of how much security information survives and how well it can be extracted, indicating that optimal usable security information can be achieved at a certain model size. We state this as the \emph{condensing law of security information}: usable security information is unimodal in model size, so security generalization error is U-shaped and is minimized at a finite $N^{\star}$ rather than decreasing without bound. The experiments on various real-world LLMs validate the condensing law, indicating that there is an optimal model size to solve security issues.
\end{abstract}


\section{Introduction}

Artificial intelligence, and in particular the large language model (LLM), has rapidly advanced in recent years. Now it plays a substantial role in domains where public welfare depends: health, transport, finance, and etc. From the release of ChatGPT in 2022, which marked the arrival of LLMs in public view, to present-day that LLMs can locate security vulnerabilities and solve open problems in mathematics, the capability of these models is no longer confined to language, they now act at the leading edge of the physical world.

The prevailing view is that, given enough training data, the final performance of
an LLM improves steadily with model size. That view produced the scaling
laws~\cite{kaplan2020scaling,hoffmann2022chinchilla}, which state that
the generalization error is a function of both data volume and model size:
\begin{equation}
L(N; D) \;=\; E \;+\; \frac{A}{N^{\alpha}} \;+\; \frac{B}{D^{\beta}} .
\label{eq:scaling}
\end{equation}
Here $E$ is the irreducible error of the language model, a property of the
intrinsic uncertainty of language itself, $N$ is the model size and $D$ is the size
of the training data. Moreover, $A$, $B$, $\alpha$ and $\beta$ are four constants capturing the linear and exponential growth coefficients in the model size and data size, respectively.

Specifically, Eq.~\eqref{eq:scaling} decomposes pre-training error into three terms: the irreducible error, the error due to limited model size, and the error that caused by limited
data. In 2022, DeepMind fitted the four unknown coefficients empirically, training over 400 models with sizes from 70~million to 16~billion parameters on corpora
from 5 to 500~billion tokens~\cite{hoffmann2022chinchilla}. With the fitted curve,
they posed a constrained optimization problem and solved for the best ratio of data size to model size under a fixed computation budget, providing a principled allocation rule for subsequent LLM training.

Scaling laws tell us that spending more data on more parameters yields a better model, and the trajectory of the field has broadly borne the prediction out: larger parameter counts have brought stronger cognitive capability.

Subsequently, as deployment matured, efficiency became a second axis of interest. Following the method of scaling laws, a group from Tsinghua University proposed the \emph{densing law}~\cite{xiao2024densing}, an empirical fit of the marginal capability gain per parameter, intended to guide the design of on-device models. It takes the form
\begin{equation}
\ln \rho_{\max} \;=\; A t + B ,
\label{eq:densing}
\end{equation}
where $\rho_{\max}$ is the capability density of a model, estimated as the ratio
between the smallest parameter count that attains the same error after
distillation and the model's actual size. By estimating the density of widely used
open-source base models released at different dates, the authors fitted $A$ and
$B$ on $29$ open-source base models, finding that capability density has risen steadily with scale, doubling approximately every $3.3$~months.

For a long time, ``larger is better'' has been the consensus. Yet an empirical regularity, however useful as guidance, may fail to explain certain domains. Security is one, and there scaling laws must be supplemented to offer real insight.

\section{Rethinking ``Larger Is Better'' in Security}
\label{sec:why}

In security, the consensus is contested from two directions at once. On the one hand, scaling laws tell us that a larger model has stronger cognitive capability and can handle more complex tasks. On the other hand, large models specialize at likelihood fitting, and are by
construction insensitive to the tail of the distribution. Yet the tail is precisely where security information lives. These two trends run against each other, so the ``larger is better'' consensus cannot be carried over to security unexamined.

Simply put, security tasks depend on information carried by anomalous data, yet anomalies are rare compared with normal traffic. Being long-tailed (e.g., zero-day attacks), they are easily overlooked by LLMs, which hurts security-task performance. More sharply, as the model size grows, their picture of the world becomes more complete and their ability to infer intent from partial input improves accordingly, as in code completion and text continuation. Reconstructing a whole from a part helps on mainstream tasks that depend on regularities of the world. On security tasks, however, it has the opposite effect: the
model uses what it knows about normal content to \emph{correct} the anomaly that carries the security signal, so that less of that signal survives into whatever downstream component has to make the security judgment.

A small model lacks the cognitive capability to extract enough task information to complete a security judgment. A large model repairs the anomalous data and so reduces the security information retained downstream. This tradeoff motivates the central conjecture of this paper.

\begin{conjecture}[Condensing law of security information]
\label{conj:main}
In the security domain, there exists an optimal model size matched to the security task, at which security information attains a condensed form.
\end{conjecture}

Section~\ref{sec:condensation} gives the functional form of the Condensing Law once the two quantities it trades off have been defined.

\section{Problem Formulation}
\label{sec:setup}

To explain why security is not a ``larger is better'' domain, and to gain a deep insight about the mechanism rather than the phenomenon, we analyze the flow of security information directly. In particular, we use the following notation.

\begin{itemize}
\item $P$: the world distribution, which is the distribution of normal data.
\item $\Pre$: the pre-training data, $\Pre \sim P$.
\item $W$: the model parameters, $W = \mathcal{F}(\Pre; U)$, where $U$ collects randomness unrelated to training.
\end{itemize}

These three variables form a Markov chain $P \rightarrow \Pre \rightarrow W$. The pre-training data is drawn from the world distribution, and the data is the
model's only channel to the world distribution.

For a test sample in the security setting, we define:

\begin{itemize}
\item $X$: the normal content of the test input, $X \sim P$; $|\Xalph|$ is the size of its alphabet (i.e. the number of values $X$ may take).
\item $\Delta$: the anomalous content, introduced by an attack (by obfuscation or by injecting an error).
\item $\Obs$: the observed input, $\Obs = \mathcal{H}(X; \Delta)$, indicating that the observed data is determined jointly by the normal and the anomalous content.
\item $S$: the security label, determined by the anomalous content, $S = \psi(\Delta)$;
      $|\Salph|$ is the size of its alphabet.
\item $Z$: the intermediate representation the model extracts.
\end{itemize}

To advance the theoretical analysis, we make two simplifying assumptions.

\begin{assumption}[Fresh sample]
\label{as:a1}
$W \perp (X, \Delta, \Obs) \mid P$, which indicates that the test sample does not appear in the
pre-training data.
\end{assumption}

\begin{assumption}[Label independent of normal content]
\label{as:a2}
$S \perp X$, indicating whether the input is anomalous is unrelated to the normal content it carries.
\end{assumption}

With this notation, we reduce the conjecture to a quantitative statement about two parameter-dependent quantities.

\begin{itemize}
\item \emph{Cognitive capability}, $\Kw = I(W;P)$: the information about the
      world distribution that the model holds.
\item \emph{Security-information retention},
      $\Rw = \frac{I(Z;S)}{I(\Obs;S)}$: the fraction of the security-relevant
      information that survives the model's feature extraction.
\end{itemize}

\section{Theoretical Analysis}
\label{sec:results}

\begin{figure}[ht]
\centering
\includegraphics[width=\columnwidth]{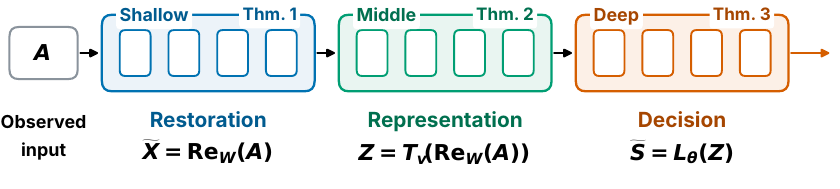}
\caption{Pipeline of LLM}
\label{fig:arichitecture}
\end{figure}

As displayed in Fig.~\ref{fig:arichitecture}, we divide the pipeline of LLM inference into three stages: (i) the restoration stage, where the shallow layers aim to repair the attacked input so that the output follows their own perception of the real-world; (ii) the representation extraction stage, where the middle layers extract features from the fundamental knowledge, which has been repaired to a familiar format, for completing downstream tasks; (iii) the decision stage, where LLMs provides the predictions by analyzing the extracted features.

Based on the decomposition, we derive the following theorems, which motivates us to propose the Condensing Law of Security Information.

\begin{theorem}[Upper bound on correction ability]
\label{thm:correct}
Let $\Xh = \mathrm{Re}_{W}(\Obs)$ be the model's estimate of the clean content
$X$ from the attacked observation $\Obs$ under parameters $W$, and let
$\varepsilon = P(X \neq \Xh)$ be its error rate. Then
\begin{equation}
1 - \varepsilon \;\le\; B_{1} + A_{1}\,\Kw ,
\label{eq:thm1}
\end{equation}
where $A_{1} = \frac{1}{\log|\Xalph|}$ and
$B_{1} = 1 - \frac{H(X \mid \Obs) - 1}{\log|\Xalph|}$.
\end{theorem}

In Theorem~\ref{thm:correct}, $1-\varepsilon$ is the probability that the model's correction restores the normal content. The larger it is, the stronger the model’s correction ability. The theorem indicates that when cognitive capability is insufficient (i.e., $\Kw$ is small), the model can correct attacked input only to a limited degree and its correction ability $1-\varepsilon$ stays small, which implies strong correction ability \emph{requires} strong cognitive capability.

For the retention of security information, we have the following theorem.

\begin{theorem}[Upper bound on retention]
\label{thm:retain}
If $Z = T_{V}\bigl[\mathrm{Re}_{W}(\Obs)\bigr]$, which indicates that $Z$ depends on $\Obs$ only
through the intermediate variable $\Xh$, then the retention of attack-related
security information satisfies
\begin{equation}
\Rw \;\le\; C_{1}\varepsilon + D_{1} ,
\label{eq:thm2}
\end{equation}
where $C_{1} = \frac{\log|\Xalph|}{I(\Obs;S)}$ and $D_{1} = \frac{1}{I(\Obs;S)}$.
\end{theorem}

Theorem~\ref{thm:retain} demonstrates that when the model corrects anomalies well (i.e., $\varepsilon$ is small), the share of security information retained in its output shrinks. 

Finally, we prove the following theorem.

\begin{figure}[t]
\centering
\includegraphics[width=0.56\columnwidth]{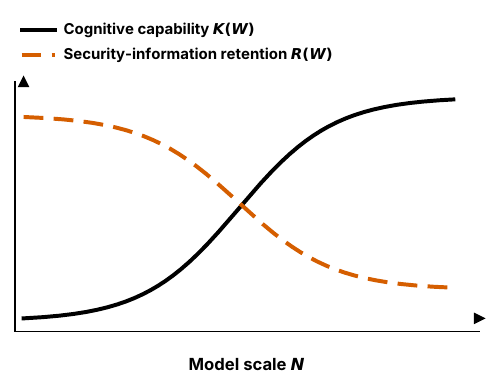}
\caption{Cognitive capability and security-information retention versus model size. Cognitive capability is upper-bounded by $H(P)$. By Theorem~\ref{thm:retain}, retention decreases affinely with capability, so against model size it is a reflected (and rescaled) copy of the capability curve.}
\label{fig:tradeoff}
\end{figure}

\begin{theorem}[Lower bound on security-task error]
\label{thm:error}
Predicting $S$ from the intermediate representation $Z$, with
$\Sh = L_{\theta}(Z)$,
\begin{equation}
P(\Sh \neq S) \;\ge\; L_{1} - M_{1}\,\Rw ,
\label{eq:thm3}
\end{equation}
where $L_{1} = \frac{H(S) - 1}{\log|\Salph|}$ and
$M_{1} = \frac{I(\Obs;S)}{\log|\Salph|}$.
\end{theorem}

Theorem~\ref{thm:error} states that as the retained security information decreases, the model’s ability to recognize attack-induced security problems weakens and the error rate rises sharply.

These theorems show that the security information retained in intermediate representations decreases as the model’s cognitive capability grows, which motivates the Condensing Law in Section~\ref{sec:condensation}.

\section{The Condensing Law}
\label{sec:condensation}

The analysis in Section~\ref{sec:results} yields the picture in Fig.~\ref{fig:tradeoff}. Specifically, scaling laws
indicate that as model size grows, the generalization error decreases and the model's knowledge of the world grows with it. Cognitive capability, in turn, is positively correlated with correction ability, and as correction ability grows, it steadily removes security information from the features passed to the higher layers.

To handle a security task efficiently, a model must be good at \emph{acquiring} security information, which depends on two things: how much security information is present at all, and how well the model can extract it. The first is
the quantity of raw material, and the second is the ability to refine it. The joint
behavior of the two curves in Fig.~\ref{fig:tradeoff} is therefore what
determines the security information a model can actually use.

\begin{figure}[h]
\centering
\includegraphics[width=0.55\columnwidth]{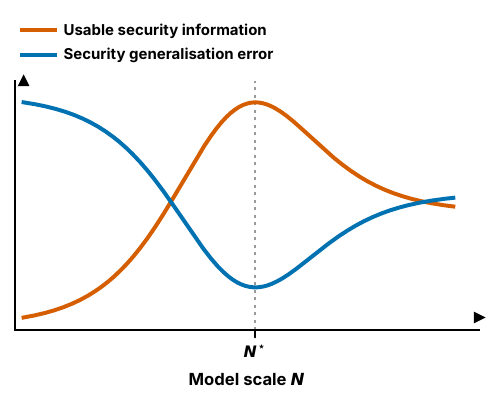}
\caption{The Condensing Law: usable security information $U(N)$ is the product of the two curves of Fig.~\ref{fig:tradeoff} and peaks at an intermediate model size
$N^{\star}$. Security generalization error is minimized at the same point.}
\label{fig:condensation}
\end{figure}

Because these two move in opposite directions with scale, their combination gives
the \emph{usable security information}
\begin{equation}
U(N) \;=\; K\bigl(W(N)\bigr) \cdot R\bigl(W(N)\bigr).
\label{eq:usable}
\end{equation}
As shown in Fig.~\ref{fig:condensation}, the product form is an ansatz, not a
consequence of Section~\ref{sec:results}. In particular, refining ability is of no use without raw material (i.e., the security information), and raw material is of no use without the ability to refine it. Composing $U$ with the relation in Theorem~\ref{thm:error} then gives a U-shaped generalization-error curve over model size.

\begin{conjecture}[Condensing law]
\label{conj:law}
$K(W(N))$ is non-decreasing in $N$ and bounded above by $H(P)$;
$R(W(N))$ is non-increasing in $N$ by Theorem~\ref{thm:retain}; and $U(N)$ in
\eqref{eq:usable} is unimodal. Consequently there is a finite
\begin{equation}
N^{\star} \;=\; \arg\max_{N} \, U(N)
\label{eq:nstar}
\end{equation}
at which the security generalization error attains its minimum, and beyond which
further scale makes security performance worse rather than better.
\end{conjecture}

Loosely speaking, as $K\bigl(W\bigr)\geq \frac{1-B_1}{A_1}-\frac{1}{A_1} \varepsilon$ and $R\bigl(W\bigr)\leq C_1 \varepsilon + D_1$, we have:

\begin{equation}\label{conjecture}
    U(N)=K\bigl(W\bigr) \cdot R\bigl(W\bigr) \approx \left[\frac{1-B_1}{A_1}-\frac{1}{A_1} \varepsilon\right] \cdot \left[ C_1 \varepsilon + D_1\right],
\end{equation}
\noindent{where} $\varepsilon$ is a function of $N$. This equation approximately holds because we cross multiply the two sides of the inequalities.

Eq.~\eqref{conjecture} is a quadratic function of $\varepsilon$, therefore has a unique maximum point in terms to $\varepsilon$. That is,

\begin{equation}
\frac{d\,U(N)}{d\,\varepsilon} = \frac{-2C_1}{A_1} \varepsilon-\frac{D_1}{A_1}+\frac{(1-B_1)C_1}{A_1}=0 \Rightarrow \varepsilon^{*} = \frac{1-B_1}{2} - \frac{D_1}{2C_1},
\end{equation}

Hence, if $\varepsilon^* = \frac{H(X|A)-2}{\log|\Xalph|}$, where $\varepsilon^* \in (0,1)$ is reasonable, the usable security information achieves its optimum 
\begin{equation}\label{eq:u_star}
    U^{*}(N)=\frac{H(X|A)^2}{4 \cdot I(A;S)}.
\end{equation}
$N^*$ is defined as the solution of Eq.~\eqref{eq:u_star}, which maximizes the usable security information to accomplish the downstream missions.

At $N^{\star}$, cognitive capability and the retention of security information are best
traded off. Hence, security information condenses there, and the model's ability to solve security problems is maximized. 

\section{Experiments}
\label{sec:experiments}

We conduct two experiments to validate the conjecture. The first tests the mechanism we propose, and the second tests the consequence the conjecture predicts.



\subsection{The Retention of Security Information Decreases with Model Size}

{\bfseries Dataset.} In the first experiment, the benign and anomalous inputs in our comparison are taken from \emph{HttpParamsDataset}~\cite{morzeux2016httpparams}, whose records are the \emph{values} of
HTTP request parameters. It holds 31{,}067 items: 19{,}304 benign and 11{,}763 anomalous, the latter split into SQL injection (10{,}852), cross-site scripting (532), path traversal (290) and command injection (89). In particular, the benign items are the payloads of the normal requests in HTTP DATASET CSIC 2010~\cite{gimenez2010csic}, automatically generated from a Spanish-language e-commerce application. The SQL injections were produced with \texttt{sqlmap}~\cite{sqlmap}, the cross-site scripting samples are generated with XSSYA and FuzzDB~\cite{fuzzdb}, and the command-injection and path-traversal samples are produced with Vega Scanner and FuzzDB.

In this experiment, we aim to validate that as the size of a model increases, the intermediate-layer representations produced by normal and anomalous inputs move closer together. This result implies that security-information retention is decreasing, validating Theorem~\ref{thm:retain}.

To achieve this target, we measure the distance between the centroids of the attacked and the normal representations at matched absolute block indices across the Pythia
suite~\cite{biderman2023pythia}, whose members share one corpus in one
order~\cite{gao2021pile} and therefore isolate scale from training-data confounds. To analyze these results in detail, we display the distances of all attacks separately
\begin{figure}[H]
\centering
\includegraphics[width=\widewidth]{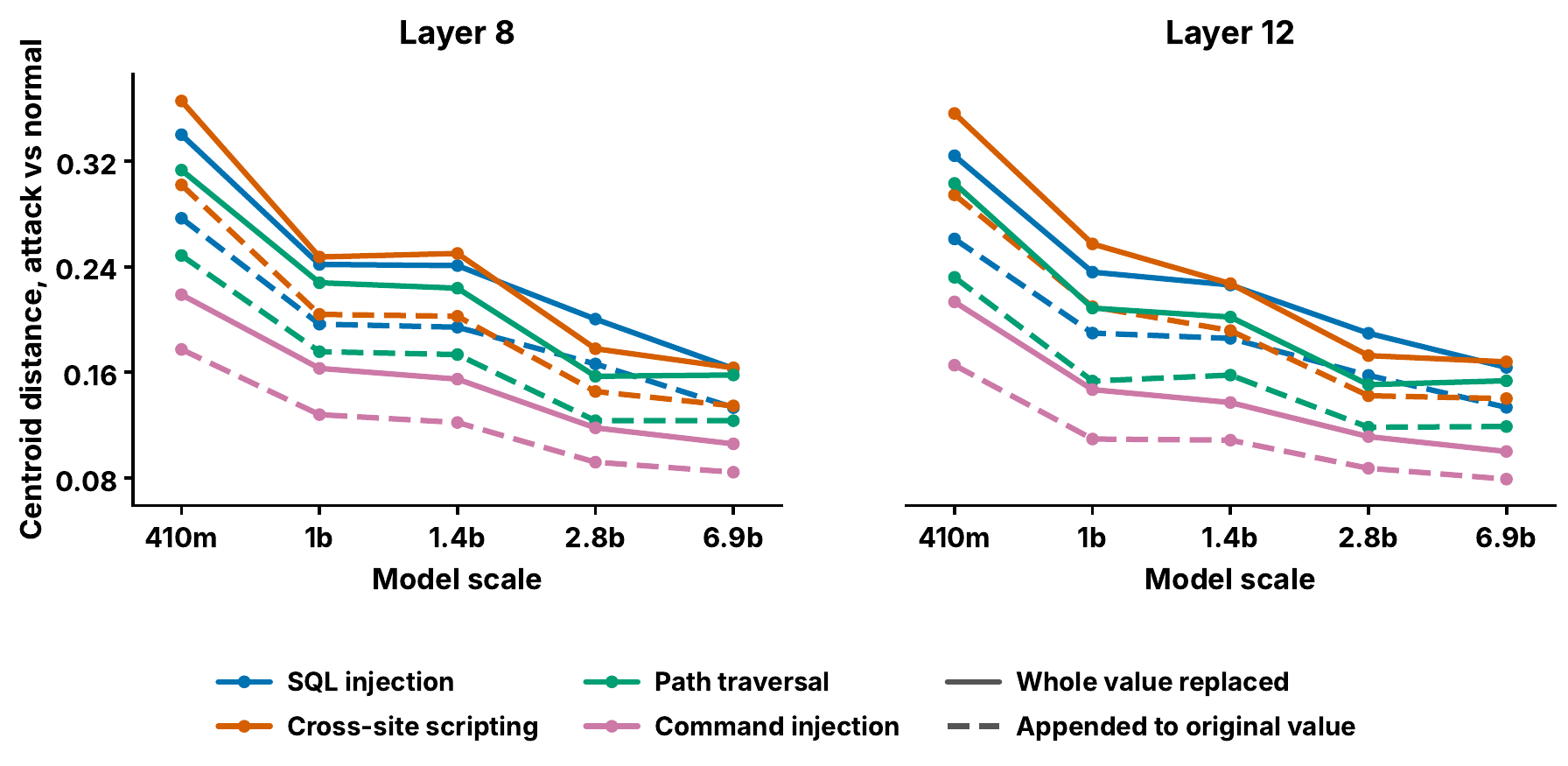}
\caption{Centroid distance between attacked and normal representations at blocks
8 and 12, with the eight attack conditions kept separate. Every condition falls
with model size.}
\label{fig:perattack}
\centering
\includegraphics[width=0.55\columnwidth]{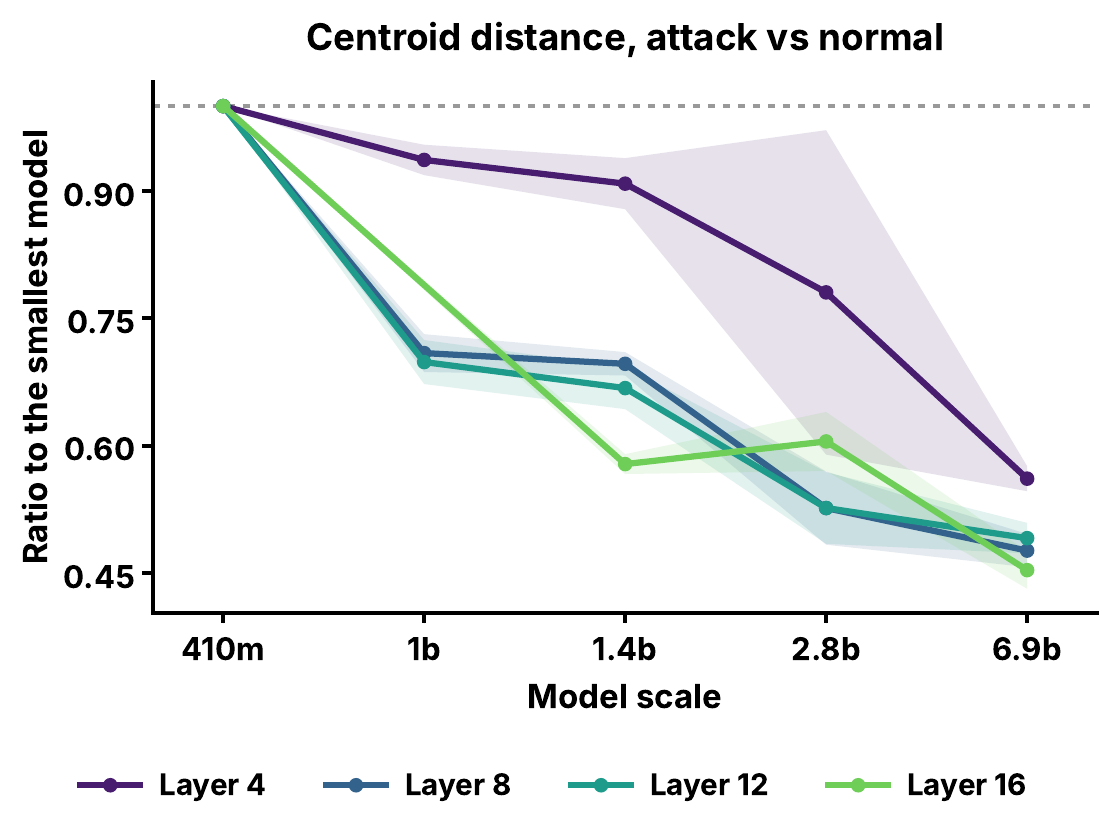}
\caption{The same measurements normalized by each condition's value at the smallest model and averaged over the eight conditions.}
\label{fig:normalised}
\end{figure}
\noindent{in} Fig.~\ref{fig:perattack}. The distance of all attacks declines with scale, which validates that the retained security information is decreasing. Moreover, Fig.~\ref{fig:normalised} normalizes each attacks by its own value at the smallest model, which removes the level differences between classes and leaves the consistency of the trend visible.

\begin{figure}[h]
\centering
\includegraphics[width=\widewidth]{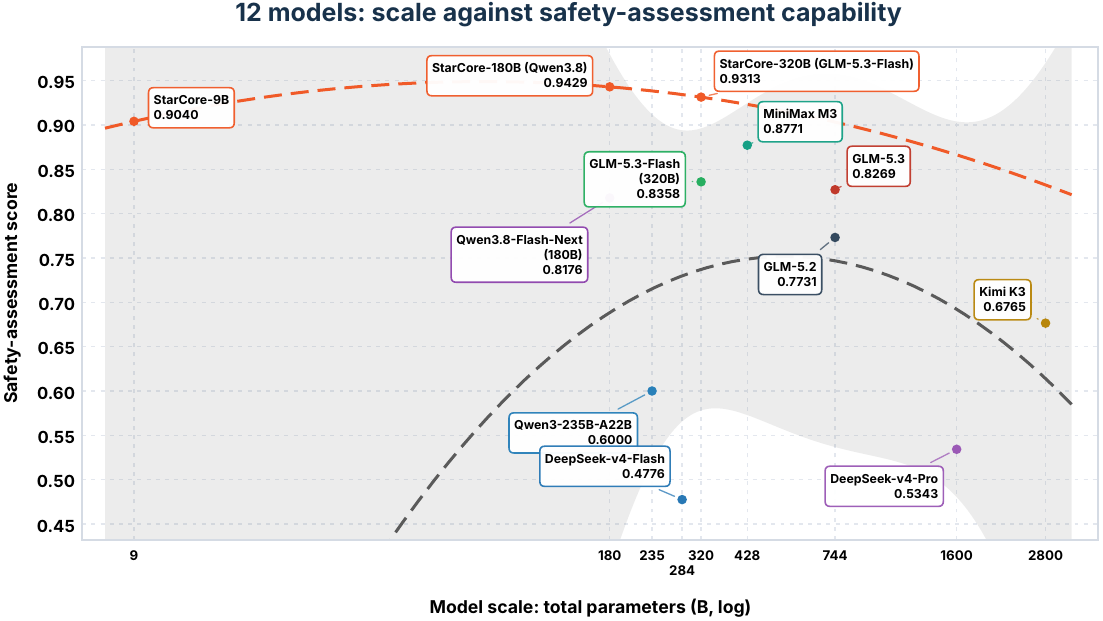}
\caption{Safety-assessment score against total parameter count for twelve models (Colors distinguish models).}
\label{fig:models12}
\end{figure}
\subsection{Real-World Fitting}

As size grows, performance on the security task has a best
fit at an intermediate point, confirming the condensing law conjecture. We use the cybersecurity world model benchmark in BigModelGym~\cite{bigmodelgym} to evaluate the real-world fitting across existing LLMs. The benchmark dataset contains 1,000 security-related data chains extracted from code (250), traffic data (250), system logs (250), and agent trajectories (250) in real-world systems.

Fig.~\ref{fig:models12} displays a safety-assessment score against total parameter scale for twelve contemporary models. In particular, the StarCore~\cite{starcore} model is fine-tuned with state chains, a new format of training data. Therefore, we fit the performance of the StarCore model and the other LLMs separately. As displayed in the figure, the performance increases as the model size is small and achieves the optimal performance at the intermediate size. After that, the performance gradually decreases. This phenomenon validates our claim of condensing law of security information. Moreover, we observe that StarCore, which even uses a smaller model size, consistently outperforms existing LLMs on security tasks, suggesting that state chains provide a more effective format for security fine-tuning.

\section{Conclusion}

Scaling laws describe a world in which more parameters lead to greater capability, and for the most part they do. Security is the domain where that capability can work against these laws, because it causes the model to repair the very anomaly on which the security decision rests. Formalizing the trade-off as one between cognitive capability $\Kw$ and security-information retention $\Rw$ gives three theorems, which predict a U-shaped security generalization error in model size. We call the resulting statement the \emph{condensing law of security information}: usable security information is unimodal in scale, so there is a finite $N^{\star}$ at which it peaks (condenses), and beyond which additional parameters reduce security performance rather than improve it. Experiments on the Pythia suite and real-world LLMs confirm our predictions: representations of normal and anomalous inputs converge as scale grows, and security-task performance peaks at an intermediate size.

\IfFileExists{IEEEtran.bst}{\bibliographystyle{IEEEtran}}{\bibliographystyle{unsrt}}
\bibliography{refs}

\appendix

\section{Proof of Theorem~\ref{thm:correct}}

By Assumption~\ref{as:a1}, $W \rightarrow P \rightarrow (X, \Obs)$ is a Markov
chain, so by the data-processing inequality~\cite{cover2006elements}
\begin{equation}
I(W; X, \Obs) \;\le\; I(W;P) \;=\; \Kw .
\end{equation}
By the chain rule,
\begin{equation}
I(W; X, \Obs) \;=\; I(W;\Obs) + I(W; X \mid \Obs) \;\ge\; I(W; X \mid \Obs).
\end{equation}
Since $\Xh = \mathrm{Re}_{W}(\Obs)$ is a function of $(\Obs, W)$, conditioning
reduces entropy and Fano's inequality~\cite{fano1961transmission} gives
\begin{equation}
H(X \mid \Obs, W) \;\le\; H(X \mid \Xh) \;\le\; \varepsilon \log|\Xalph| + 1 .
\end{equation}
From the relation between entropy and mutual information,
\begin{equation}
\begin{aligned}
I(W; X \mid \Obs) &= H(X \mid \Obs) - H(X \mid \Obs, W) \\
                  &\ge H(X \mid \Obs) - \varepsilon\log|\Xalph| - 1 .
\end{aligned}
\end{equation}
Combining the two inequalities,
\begin{equation}
\Kw \;\ge\; H(X \mid \Obs) - \varepsilon\log|\Xalph| - 1 ,
\end{equation}
and rearranging,
\begin{equation}
1 - \varepsilon \;\le\; \frac{\Kw}{\log|\Xalph|}
 - \frac{H(X \mid \Obs) - 1}{\log|\Xalph|} + 1 ,
\end{equation}
which is Eq.~\eqref{eq:thm1}. \hfill$\blacksquare$

\section{Proof of Theorem~\ref{thm:retain}}

Since $Z = T_{V}(\Xh)$, all security information in $Z$ arrives through $\Xh$, and
$S \leftarrow \Delta \rightarrow Z$ is a fork. By the data-processing
inequality,
\begin{equation}
I(Z;S) \;\le\; I(\Xh;S) .
\end{equation}
For $I(\Xh;S)$,
\begin{equation}
I(\Xh;S) \;\le\; I(\Xh, X; S) \;=\; I(X;S) + I(\Xh; S \mid X).
\end{equation}
By Assumption~\ref{as:a2}, $I(X;S) = 0$, so
\begin{equation}
I(\Xh;S) \;\le\; I(\Xh; S \mid X) \;\le\; H(\Xh \mid X).
\end{equation}
Since $\varepsilon = P(X \neq \Xh)$, Fano's inequality applied with the roles of
$X$ and $\Xh$ exchanged gives
\begin{equation}
H(\Xh \mid X) \;\le\; \varepsilon\log|\Xalph| + 1 .
\end{equation}
Therefore
\begin{equation}
\Rw \;=\; \frac{I(Z;S)}{I(\Obs;S)}
 \;\le\; \frac{\log|\Xalph|}{I(\Obs;S)}\,\varepsilon + \frac{1}{I(\Obs;S)} ,
\end{equation}
which is Eq.~\eqref{eq:thm2}. \hfill$\blacksquare$

\section{Proof of Theorem~\ref{thm:error}}

By Fano's inequality,
\begin{equation}
P(\Sh \neq S) \;\ge\; \frac{H(S \mid Z) - 1}{\log|\Salph|} .
\end{equation}
Since $H(S \mid Z) = H(S) - I(S;Z)$ and $I(S;Z) = \Rw\,I(\Obs;S)$,
\begin{equation}
P(\Sh \neq S) \;\ge\; \frac{H(S) - 1}{\log|\Salph|}
 - \frac{I(\Obs;S)}{\log|\Salph|}\,\Rw ,
\end{equation}
which is Eq.~\eqref{eq:thm3}. \hfill$\blacksquare$

\end{document}